\documentclass[fleqn,10pt]{wlscirep}
\usepackage[utf8]{inputenc}
\usepackage[T1]{fontenc}
\usepackage{caption}
\usepackage{subcaption}
\usepackage{comment}
\usepackage{xcolor}
\title{Evaluating Efficacy of Indoor Non-Pharmaceutical Interventions against COVID-19 Outbreaks with a Coupled Spatial-SIR Agent-Based Simulation Framework.}

\author[1,2,*]{Chathika Gunaratne}
\author[1]{Rene Reyes} 
\author[1]{Erik Hemberg}
\author[1]{Una-May O'Reilly}

\affil[1]{Massachusetts Institute of Technology, Computer Science and Artificial Intelligence Laboratory, Cambridge, MA, USA}
\affil[2]{Current Affiliation: Oak Ridge National Laboratory, Oak Ridge, TN, USA}

\affil[*]{chathika@mit.edu}

\begin{abstract}
    Contagious respiratory diseases, such as COVID-19, depend on sufficiently prolonged exposures for the successful transmission of the underlying pathogen. It is important for organizations to evaluate the efficacy of interventions aiming at mitigating viral transmission among their personnel. We have developed a operational risk assessment simulation framework that couples a spatial agent-based model of movement with a SIR epidemiological model to assess the relative risks of different intervention strategies. By applying our model on MIT's STATA building, we assess the impacts of three possible dimensions of intervention: one-way vs unrestricted movement, population size allowed onsite, and frequency of leaving designated work location for breaks. We find that there is no significant impact made by one-way movement restrictions over unrestricted movement. Instead, we find that a combination of lowering the number of individuals admitted below the current recommendations and advising individuals to reduce the frequency at which they leave their workstations lowers the likelihood of highly connected individuals within the contact networks that emerge, which in turn lowers the overall risk of infection. We discover three classes of possible interventions based on their epidemiological effects. By assuming a direct relationship between data on secondary attack rates and transmissibility in the SIR model, we compare relative infection risk of four respiratory diseases, MERS, SARS, COVID-19, and Measles, within the simulated area, and recommend appropriate intervention guidelines. 
\end{abstract}

\begin{document}

\flushbottom
\maketitle
\thispagestyle{empty}


\section*{Introduction}

Establishing safe return-to-work guidelines is essential to guaranteeing COVID-19 outbreaks. Respiratory diseases such as COVID-19 are often spread through droplet or aerosol transmission of the virus \cite{hertzberg2018behaviors,jarvis2020aerosol,kohanski2020review,smith2020aerosol}, where spatial proximity and duration of closeness are important factors towards the transmission of pathogens. Therefore, it is vital that organizations properly assesses the spatial effects of imposed safety measures on their ability to reduce risk of infection among the population by suppressing prolonged contacts. Studies have shown that the presence of highly connected individuals, or hubs, in social networks, is a leading factor causing large outbreaks among a population \cite{kojaku2021effectiveness}. Agent-based models are ideal at replicating real-world spatial movement patterns and have been used to assess the spread of infectious diseases such as COVID-19 in both indoor and outdoor settings \cite{bedson2021review,cliff2018investigating,gharakhanlou2020spatio,zhou2021optimizing,hackl2019epidemic,hunter2017taxonomy,abdulkareem2018intelligent}. Floor layouts, walls, hallways, and other physical obstacles, in addition to safety guidelines such as recommended break durations, may restrict certain contacts from occurring while amplifying others, and spatial agent-based models are able to simulate such effects. Modeling these spatial effects provides a more accurate representation of the plausible patterns of prolonged exposures that may occur within a workplace and help in the generation of contact networks that can then be studied on their resilience to disease outbreaks.

We demonstrate how spatial agent-based modeling can be used to predetermine contact networks under varying intervention strategies, under constraints specified by the CDC definition of a \textit{prolonged contact} for SARS-CoV-2 transmission \cite{CDCprevention_2021}. We generated prolonged-contact networks using a spatial-agent based model implemented in AnyLogic, which incorporates room and corridor locations, entrances and exits, arrival schedules, restroom locations, and break areas. 
We investigate the effects of intervention strategies based on three dimensions, forms of movement restriction, population size, and frequency of leaving designated office for breaks. The generated networks were then subjected to an agent-based SIR model to compare the resilience of the simulated intervention strategies against diseases with different transmissibilities. Finally, we used known secondary attack rates for four respiratory diseases, MERS, SARS, COVID-19, and Measles, in order to approximate final infection ratios expected for each disease under the simulated intervention strategies. 

\section*{Methodology}
\begin{figure}[ht]
    \centering
    \includegraphics[width=0.8\linewidth]{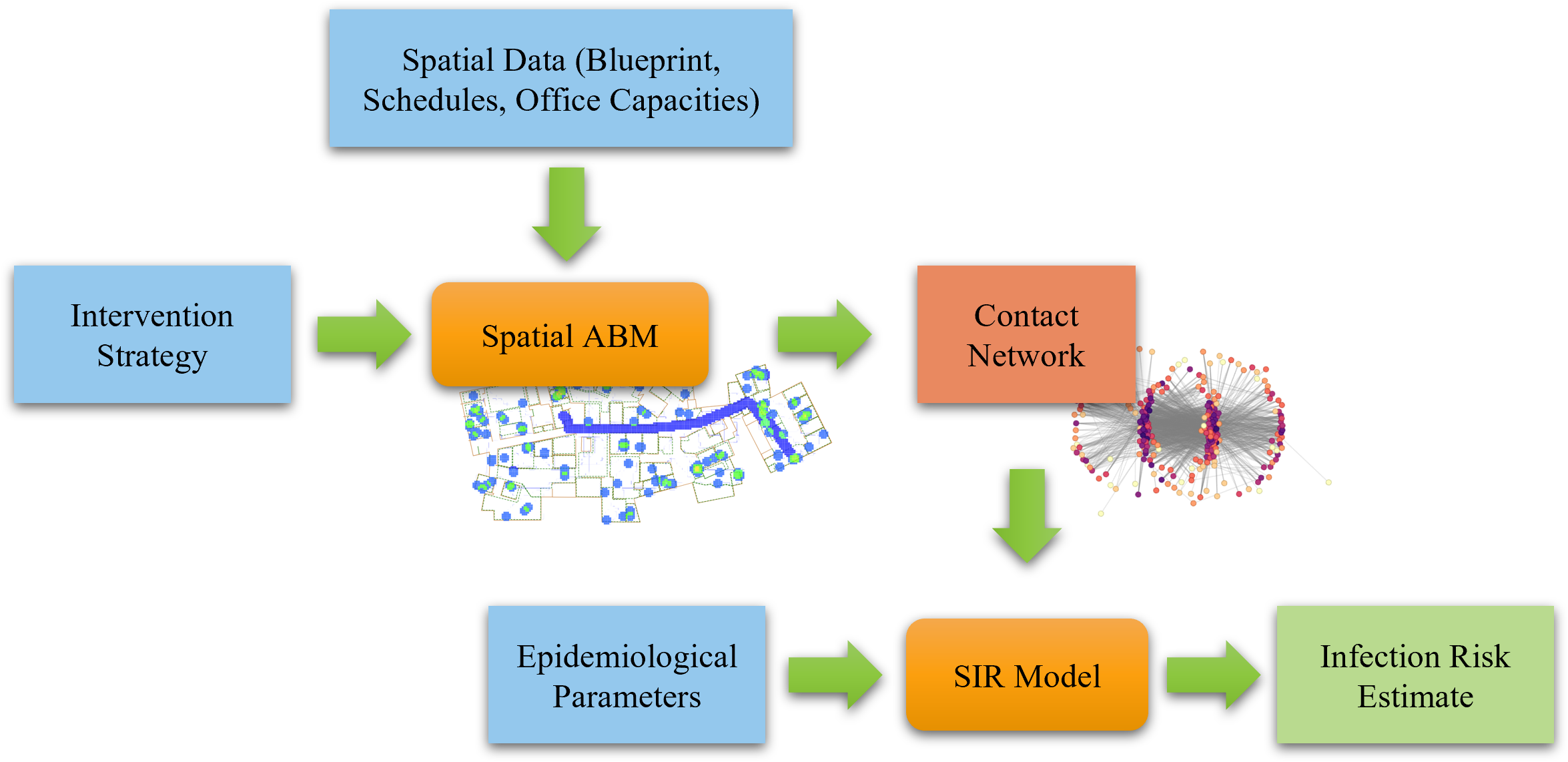}
    \caption{System diagram}
    \label{fig:SystemDiagram}
\end{figure}

We coupled two agent-based models to assess operational efficacy of COVID-19 intervention strategies on the STATA building at MIT by assessing resulting risks of infection. First, a data-informed spatial agent-based model was used to simulate individual movement within a floor of the building under varying intervention constraints, to generate high-risk contact networks. Second, a SIR model was simulated on the resulting high-risk contact networks to obtain final infection ratios of the population. Fig. \ref{fig:SystemDiagram}.

\subsection*{Spatial Agent-Based Model}
The spatial agent-based model was implemented to mimic the movement of personnel under restrictions imposed by constraints of the varying intervention strategies. The model was implemented with the AnyLogic simulation software, which allowed for accurate representation of physical spaces including offices, stairwells, and restrooms within the simulated area. The second floor of the STATA building was used for this purpose, which hosts several CSAIL research offices and labs, conference rooms, rest areas, two restrooms, utility space, and pantry area. The blueprint image of the building floor under consideration was loaded into AnyLogic and AnyLogic's \textit{wall} construct from the \textit{pedestrian library} was used to designate areas impassable by agents, forming office spaces and rooms. Stairwells and elevators were identified from the blueprints and \textit{targetlines} from the \textit{pedestrian library} were allocated at these points within the model from which agents could enter or exit the floor. Office spaces were demarcated with their real-world identifiers and assigned capacities according to the current office capacities set by building administration. 

\begin{figure}[ht]
    \centering
    \includegraphics[width=0.5\linewidth]{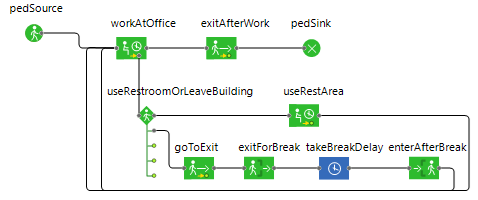}
    \caption{State machine driving agent behavior within the simulated environment of the spatial agent-based model, implemented using AnyLogic's Pedestrian library.}
    \label{fig:SpatialModelStateMachine}
\end{figure}

\begin{figure}[ht]
    \centering
    \includegraphics[width=\linewidth]{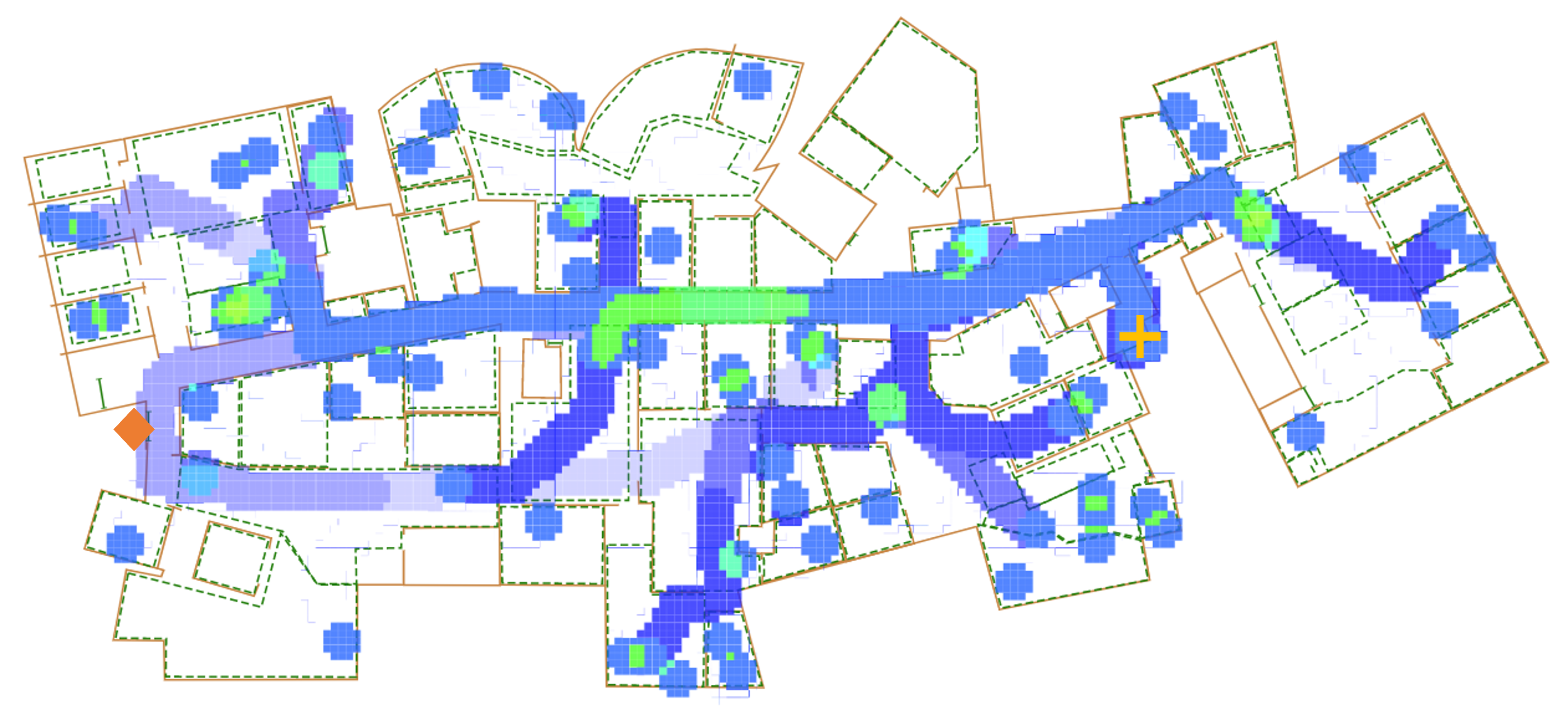}
    \caption{Example run of the spatial agent-based model with one-way movement restriction enabled. A heatmap depicting agent movement through the simulated floor is overlayed (lighter colors showing where agents have recently spent more time at). The designated entrance is marked with a yellow plus symbol and the designated exist is marked with an orange diamond symbol. In unrestricted movement, both these locations can be used for both entrance and exit to the space.}
    \label{fig:Model_Example}
\end{figure}

Agent behavior was controlled using a state machine implemented with components from AnyLogic's \textit{pedestrian library} as shown in Fig. \ref{fig:SpatialModelStateMachine}. Individuals were generated from a pedestrian source, at which they were randomly assigned to an available office. We used an approximation of a daily schedule for the floors as shown in Tab. \ref{tab:Schedule}. Individuals began entering the floor from around 6am at a very low rate, followed by a gradual increase in rate towards 9am, ending with a gradual decrease in rate towards 11am. We assumed that individuals would work a complete 8 hour shift after which they would exit the floor. This meant that departure times would follow a symmetrical distribution to that of entrance times. The agent would then enter the simulation from a designated stairwell or elevator according to the intervention strategy's movement restriction as detailed below. Once on the floor, the agent would move towards its office location along the hallways created by the wall objects. Collision detection and path-finding algorithms were handled by AnyLogic and comfortable walking speed was selected for each individual from a uniform distribution between 0.5 and 1 meters per second. Once at the office, individuals remained at a chosen location within the office space, unless they were interrupted to take a break at an hourly probability of $\alpha$. There was a 0.5 probability that they would remain on the floor in designated break locations and a 0.5 probability that they would exit the floor during this time. Break locations included the restrooms and common areas on the floor and the time agents would spend at a break location followed a uniform distribution between 5 and 20 minutes. Individuals that exit the floor during their break would use designated exits and entrances according to the movement restriction in place and the time spent outside was chosen from a triangular distribution of minimum, maximum, and mode of 20, 60, and 30 minutes, respectively. 
\begin{table}[ht]
    \centering
    \begin{tabular}{c|c|c}
         \textbf{Start Time} & \textbf{End Time} & \textbf{Proportion of Population} \\
         \hline
         6:00am & 7:00am & 0.05\\
         7:00am & 8:00am & 0.15\\
         8:00am & 9:00am & 0.4\\
         9:00am & 10:00am & 0.3\\
         10:00am & 11:00am & 0.1
    \end{tabular}
    \caption{Schedule used for controlling agent entrance to the floor over time. Each hour, the specified proportion of the population enters the floor from the designated entrances and heads to their respective office spaces. Agents spend 8 hours within the simulation and then exit for the day, implicitly making the exit schedule symmetrically distributed.}
    \label{tab:Schedule}
\end{table}
Intervention strategies were decomposed into three dimensions, namely movement restriction, population capacity, hourly break probability. Two movement restrictions were simulated: 1) unrestricted movement, where individuals were able to enter and exit the floor from any of the available stairwells and elevators, and 2) one-way movement, where entrance and exits to the floor were only allowed at separate, designated locations as shown in the map in Fig. \ref{fig:Model_Example}. Population capacities were controlled using a population multiplier parameter, $\beta$, applied to the recommended capacities of each office. For each office, $o$, the population capacity, $n_o$ was calculated as $n_o = \beta c_o$, where $c_o$ was the capacity for $o$ recommended by the administration. Thus the total population for the simulation was $\sum_{o \in O}{\beta c_o}$, where $O$ is the complete collection of offices. Hourly probability of taking a break, $\alpha$, determined whether agents would leave their offices for a break as described above and utilize the break locations or exit the floor temporarily, controlling the rate at which agents may contact one another in the hallways or common areas. $\beta$ was varied in the range $[0.25, 2.0]$ in increments of $0.25$ and $\alpha$ was varied in the range $[0.05, 0.45]$ in increments of $0.05$, and along with the two forms of movement restriction resulted in 128 separate strategies. Each strategy simulation was replicated 10 time to account for stochasticity. 

Contacts leading to high infection risk were measured from simulations of the spatial ABM by considering CDC definition of a prolonged contact between an infected individual and a susceptible individual leading to a high risk of transmission of the SARS-CoV-2 virus \cite{CDCprevention_2021}. According to CDC guidelines, a prolonged contact occurs when a susceptible individual and infected individual come within 6 feet of one another for over a 15 minute duration with minimal PPE usage. We used this definition to track agents in our model that came into prolonged contact by recording instances in the spatial ABM simulations when two agents spend 15 minutes or more within 6 feet of one another. These contacts are are edges to construct an undirected prolonged contact network for each simulation.

\subsection*{Epidemic Simulation on Prolonged Contact Networks}
An SIR model was used to simulate the viral spread over the contact networks generated from the spatial ABM. The goal of the SIR model was to assess the infection risk posed by the intervention strategies that produced the contact networks being simulated, when a single infected individual would arrive during a workday. The SIR model was executed as follows on each contact network. We assume a parameter $\rho$ to represent the transmissability of the disease, where $\rho$ equals the proportion of susceptible neighbors of an infected agent that will be infected. Say, that agent state is represented by the function $\theta : M -> \{S,I,R\}$, where $M$ is the set of all agents in the contact network and $\{S,I,R\}$ are the susceptible, infectious, and recovered states that any agent $m \in M$ may exist in. All agents but one were set to a susceptible state ($\theta(m_s)=S \ni m_s \in M$), except a single agent $m_i \in M$ that was chosen at random from the contact network and set to an infectious state ($\theta(m_i)=I$). A proportion, $\rho$, of the neighbors of $m$, $N_m \ni N_m \subset M, |N_m| = \rho |M| $, was then chosen and changed to infectious ($\theta(N_m)=I$). The infecting agent was then changed to a recovered state ($\theta(m_i)=R$) and could no longer infect other agents. The process was then repeated for the newly infected agents and continued until no further infected agents remained ($\theta(M) = \emptyset$). Each contact network and $\rho$ configuration, was replicated 10 times to account for stochasticity and heterogeneity in node degree centrality. The final infected ratio of the entire population, $\phi$, was measured from each SIR simulation as the proportion of recovered agents $\Phi$ at the end  of the simulation: 
\begin{align}
\label{eq:populationSIR}
\phi =& \frac{|\{\Phi | \Phi \subset M, \theta(\Phi)= R\}|}{\sum_{o \in O}{\beta c_o}}
\end{align}

Finally, we use secondary attack rate (SAR) in order to assess the efficacy of the simulated interventions at minimizing $\phi$ for the respiratory diseases, MERS, SARS, COVID-19, and Measles. SAR, also known as secondary infection risk, is an epidemiological measurement of the proportion of susceptible individuals that are infected due to close contact with an infected individual \cite{leung2021transmissibility}. SAR is especially useful for quantifying \textit{household infectivness} as it considers contacts that occur among individuals within close confines, and is applicable to our study, where individuals spend most of their workday in the same shared space. 
Assuming SAR to be a good approximation of $\rho$ in our model, we use empirically estimated 95\% confidence intervals of mean SAR for the four diseases from the literature\cite{madewell2020household, leung2021transmissibility} to define ranges of $\rho$ as follows, MERS: $[0.009, 0.107]$, SARS $[0.048, 0.107]$, COVID-19 $[0.14, 0.20]$, and Measles $[0.520, 0.846]$. Thus, by running the SIR model with $\rho$ within the respective range for the simulated disease, we are able to provide predicted final infection ratios for each disease under the simulated intervention strategies.

\section*{Results}
\begin{figure}[ht]
     \centering
     \begin{subfigure}[b]{0.45\linewidth}
         \centering
         \includegraphics[width=\linewidth]{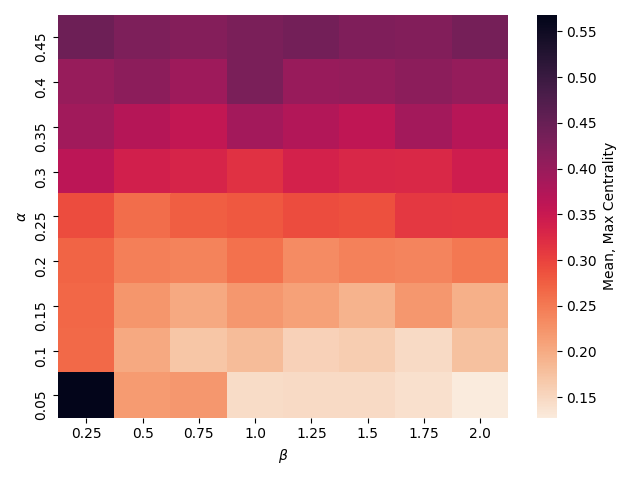}
         \caption{One-way movement}
         \label{fig:alpha_beta_centrality_a}
     \end{subfigure}
     \begin{subfigure}[b]{0.45\linewidth}
         \centering
         \includegraphics[width=\linewidth]{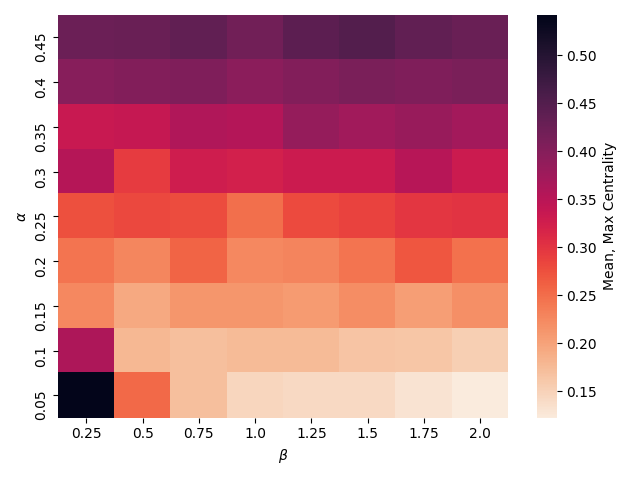}
         \caption{Unrestricted movement}
         \label{fig:alpha_beta_centrality_b}
     \end{subfigure}
        \caption{Mean maximum centrality, $\overline{C_{\textnormal{max}}}$, of prolonged-contact networks generated under varying $\alpha$, $\beta$, and form of movement restriction.}
        \label{fig:alpha_beta_centrality}
\end{figure}

Fig. \ref{fig:alpha_beta_centrality} displays, $\overline{C_{\textnormal{max}}}$, the mean of maximum degree centralities of the prolonged-contact networks generated by the spatial ABM for each intervention strategy. $\alpha$ has the greatest effect on $\overline{C_{\textnormal{max}}}$, with $\alpha=0.45$ being sufficient to produce at least one \textit{hub}, connected to nearly half of the network. There is no change in $\overline{C_{\textnormal{max}}}$ with movement restriction (Figure~\ref{fig:alpha_beta_centrality_a} and \ref{fig:alpha_beta_centrality_b}). $\alpha$ is seen to have a significantly positive correlation with $\overline{C_{\textnormal{max}}}$ (Spearman rank correlation: $r_s=0.8674$, $\textrm{p-value}=0.0$) and a significant, yet weaker, positive correlation is seen between $\beta$ and $\overline{C_{\textnormal{max}}}$ (Spearman rank correlation: $r_s=0.08771$, $\textrm{p-value}=1.7781\times10^{-195}$).

\begin{figure}[ht]
    \centering
    \includegraphics[width=\linewidth]{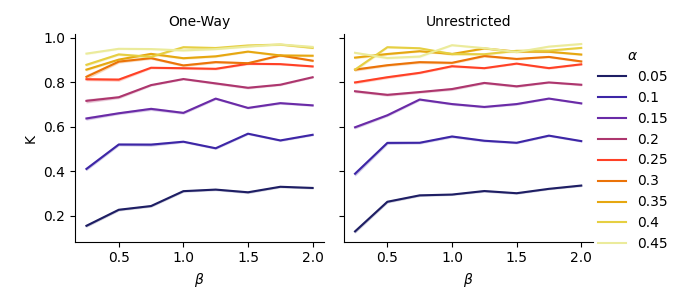}
    \caption{Proportion of agents with at least one prolonged contact $K$ under varying $\alpha$, $\beta$, and form of movement restriction.}
    \label{fig:at_least_one}
\end{figure}
Fig. \ref{fig:at_least_one} displays the proportion of agents in each simulation with at least one contact, $K$, against the three dimensions of intervention. Since the total population of each simulation was known, $K$ was calculated as: 
\begin{align}
\label{eq:contact}
K =& \frac{|M|} {\sum_{o \in O}{\beta c_o}}
\end{align}
Again, it is seen that $\alpha$ has the strongest effect on $K$, (Spearman rank correlation: $r=0.8952$, $\textrm{p-value}=0.0$). When $\alpha > 0.1$, at least more than half the population was expected to have at least one prolonged contact. $\beta$ had a slight positive correlation with $K$ (Spearman rank correlation: $r=0.0989$, $\textrm{p-value}=0.0$) , which diminished at higher values of $\alpha$. Once more, the two forms of movement have restriction have no apparent effect on $K$.

\begin{figure}[ht]
    \centering
    \includegraphics[width=\linewidth]{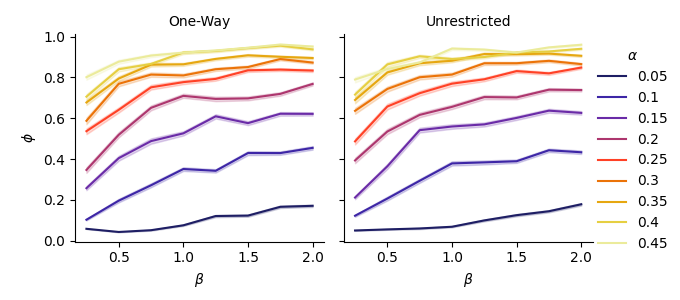}
    \caption{Final ratio infected by the number of times larger the population is by recommended capacity, by hourly break probability, for one-way and unrestricted movement.}
    \label{fig:PopMulvsInfRatiovsStrategy}
\end{figure}

Fig. \ref{fig:PopMulvsInfRatiovsStrategy}, displays outbreak size as final infected ratio, $\phi$, by $\beta$ and $\alpha$ for both forms of movement. Both forms of movement show similar behaviors under varying $\alpha$ and $\beta$. There is a significant positive correlation between $\alpha$ and $\phi$ (Spearman rank correlation: $r = 0.7954$, $\textrm{p-value}=0.0$) and a weaker positive correlation between $\beta$ and $\phi$ (Spearman rank correlation $r=0.1916$, $\textrm{p-value}=0.0$). At the recommended population capacity ($\beta = 1$), $\phi$ can be maintained under $0.5$, if $\alpha < 0.1$, while more than half the population is at risk when $\alpha > 0.1$. At $\alpha = 0.05$, it is possible to keep the final infection ratio below a complete contagion even when double the recommended population capacity is present, i.e. $\beta=2$. 

\begin{figure}[ht]
    \centering
    \includegraphics[width=\linewidth]{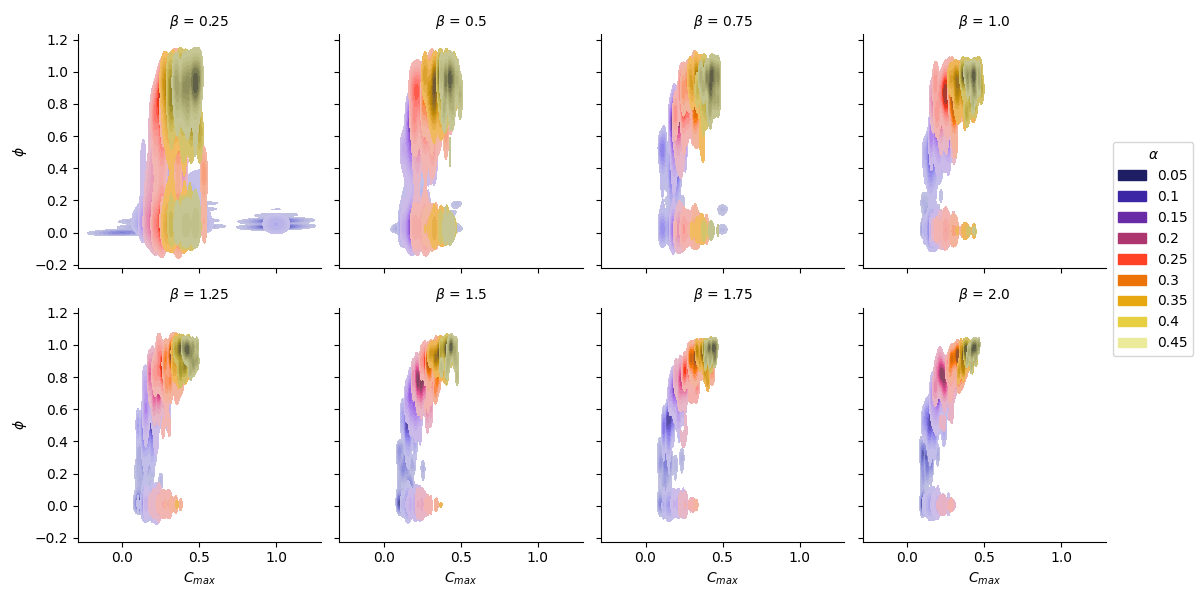}
    \caption{Density plot of $\phi$ by maximal network centrality, $C_\textnormal{max}$, under varying $\alpha$ and $\beta$.}
    \label{fig:CentralityAlphaBetaPhi}
\end{figure}

Fig. \ref{fig:CentralityAlphaBetaPhi} displays the relationship between maximum degree centrality per contact network, $C_{\textnormal{max}}$, and $\phi$, for varying $\alpha$ and $\beta$. The correlation between $\alpha$ and $C_{\textnormal{max}}$ is intensified with increasing $\beta$. At high $\beta$, a non-linear relationship between $C_{\textnormal{max}}$ and $\phi$ can be seen. In other words, for sufficiently large population sizes, higher hourly break probability can cause networks with higher maximum centrality, leading to larger networks with highly connected hubs that allow for higher final infection ratios.

\begin{figure}[ht]
\centering
\rotatebox{0}{
    \includegraphics[height=0.5\linewidth]{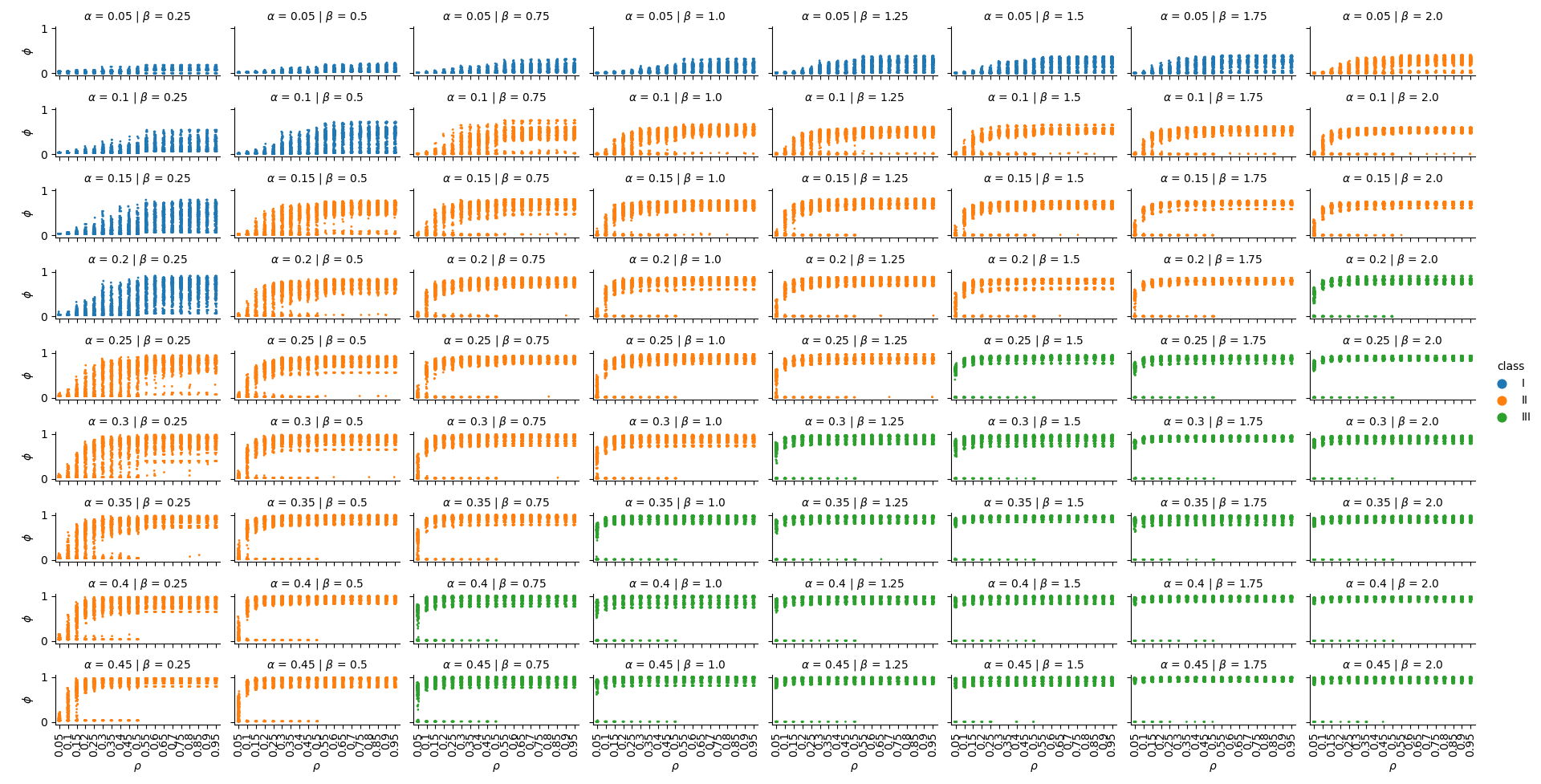}
    }
    \caption{The effect of $\rho$ on $\phi$ under different $\alpha$ and $\beta$. Three classes of behavior can be seen.}
    \label{fig:PopMul_ProbBreak_Rho_InfRatio}
\end{figure}

Fig. \ref{fig:PopMul_ProbBreak_Rho_InfRatio}, shows the effects of $\rho$ on $\phi$ under different $\alpha$ and $\beta$ conditions. Under varying $\alpha$ and $\beta$, the effect that $\rho$ has on $\phi$ can be separated into thee classes. Class I (blue) where gradually larger ranges of $\phi$ can be expected with higher $\rho$, but $\phi$ stays generally less than 1; Class II (orange) where there is an initial gradual increase in possible range of $\phi$ with $\rho$, followed by a bifurcation into an oscillation of period 2, i.e. either very low $\phi$ or higher ranges of $\phi$ ($> 0.5$), which eventually transitions into $\phi \approx 1$ for $\rho > 0.5$; and Class III (green) where the final states exist in an oscillation of period 2 for $\rho <= 0.5$, at $\phi > 0.75$ and $\phi \approx 0$, and for $\rho > 0.5$ only $\phi > 0.75$ is observed.

\begin{figure}[ht]
    \centering
    \includegraphics[width=\linewidth]{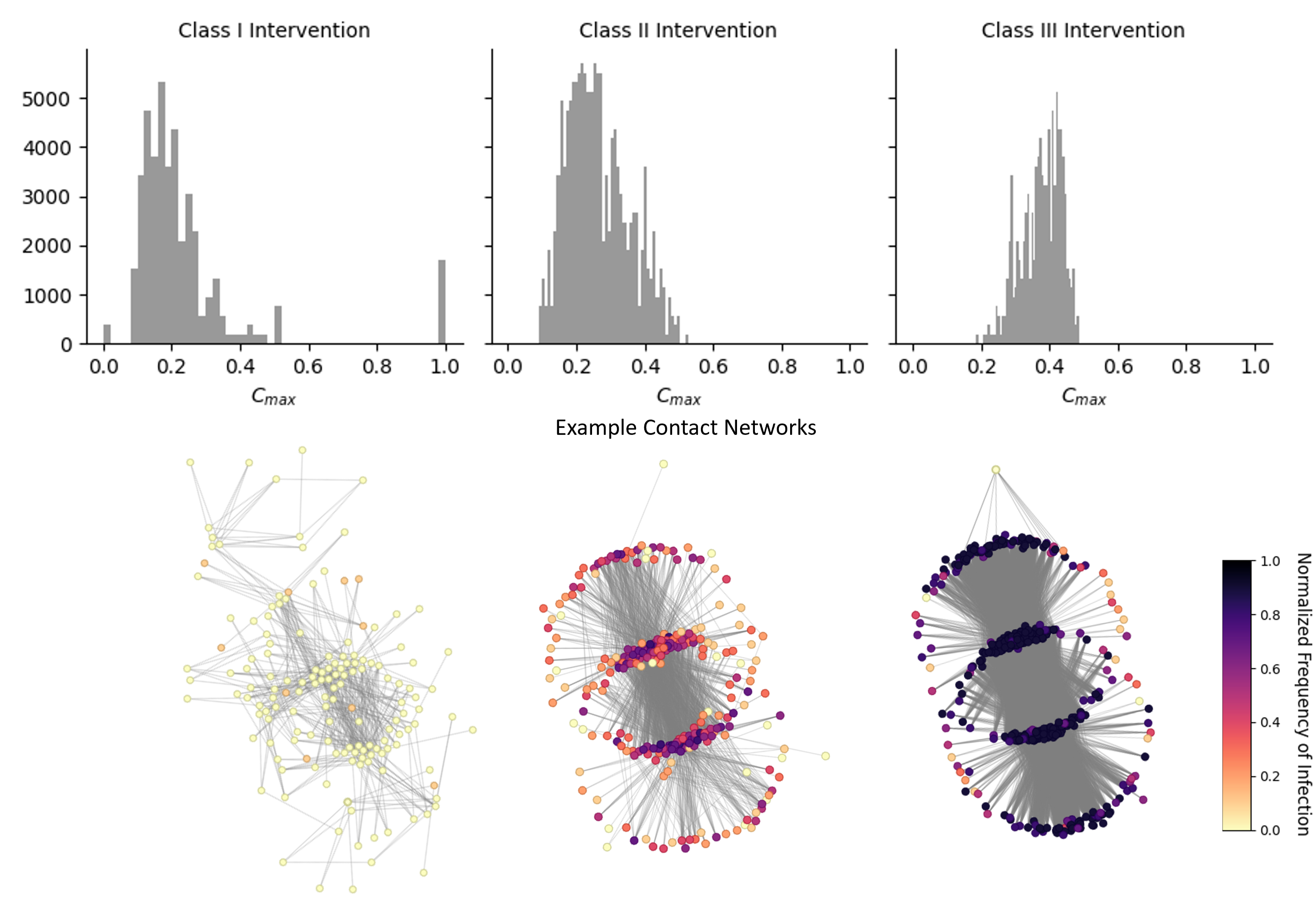}
    \caption{Maximum centrality distributions of contact networks for each intervention class, along with sample networks from all three classes. Colors of each node in the sample networks depict the normalized frequency of infection per node over 10 SIR simulation on the example network (darker colors represent higher risk).}
    \label{fig:Class_Centrality_SampleNet}
\end{figure}

Fig. \ref{fig:Class_Centrality_SampleNet} displays $C_{\textnormal{max}}$ distributions and example networks for all three classes of intervention. The normalized frequency of infection for each node in the example networks, over all simulations, is also shown. It can be seen that Class III networks tend to have hubs that are connected to between $20\%$ to $50\%$ of the network. This number is less for Class II and least for Class I. Furthermore, the example networks show how these hubs in Class III networks are nearly always likely to be infected, which in turn exposes the many nodes they are connected to (nearly half of the network). This effect is less in Class II, while in Class I the hubs have approximately the same (or even less) likelihood of infection in comparisons to other nodes.

\begin{figure}[ht]
    \centering
    \includegraphics[width=\linewidth]{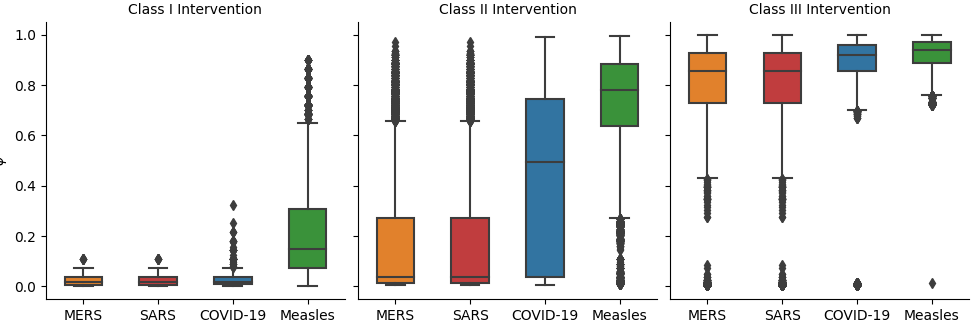}
    \caption{Effectiveness of Class I, II, and III strategies on outbreak size measured as final infected ratio, $\phi$, for different contagious respiratory diseases compared by secondary attack rate.}
    \label{fig:Class_Virus_Phi}
\end{figure}

Finally, we compare the predicted efficacy of Class I, II, and III interventions on $\phi$ by relating $\rho$ to SAR for MERS, SARS, COVID-19, and Measles. Fig. \ref{fig:Class_Virus_Phi} displays the predicted $\phi$ these diseases would have in the simulated environment under the three classes of intervention. The strictest Class I interventions are likely to prevent large outbreaks of MERS, SARS, and COVID-19, yet, Measles being an extremely transmissible disease, can result in moderate outbreaks even under the strictest conditions. Class II interventions are likely to prevent MERS and SARS outbreaks, but are ineffective against Measles, and likely to allow significant outbreaks of COVID-19. Class III interventions are likely to allow significantly large outbreaks of all four diseases with nearly the entire population at risk of infection. For each disease, Mann Whitney U tests were performed to confirm whether there was a statistically significant improvement by having Class I interventions over Class II interventions ($H_0 : \textrm{Class I} \: \phi >= \textrm{Class II} \: \phi$) and Class II interventions over Class III interventions ($H_0 : \textrm{Class II} \: \phi >= \textrm{Class III}  \: \phi$). The results for the Mann Whitney U tests, shown in tab. \ref{tab:class_comparison} show that in all four cases, Class I interventions were likely to have a significant reduction in final infected ratios over Class II interventions, as did Class II interventions over Class III interventions.

\begin{table}[]
    \centering
    \begin{tabular}{r|c|c}
    \toprule
    Disease & $H_0 : \textrm{Class I} \: \phi >= \textrm{Class II} \: \phi$ & $H_0 : \textrm{Class II} \: \phi >= \textrm{Class III}  \: \phi$\\\midrule
        MERS & ($U=22359759.5$, $\textrm{p-value}=2.673\times10^{-194}$) & ($U=19540696.5$, $\textrm{p-value}=0.0$)\\
        SARS & ($U=22359759.5$, $\textrm{p-value}=2.673\times10^{-194}$) &($U=19540696.5$, $\textrm{p-value}=0.0$)\\
        COVID-19 & ($U=3048844.0$, $\textrm{p-value}=0.0$) &($U=3563600.5$, $\textrm{p-value}=0.0$)\\
        Measles & ($U=26869646.0$, $\textrm{p-value}=0.0$) &$U=202887512.0$, $\textrm{p-value}=0.0$)\\\bottomrule
    \end{tabular}
    \caption{Results from Mann-Whitney U tests for the null hypotheses $ H_0 : \textrm{Class I} \: \phi >= \textrm{Class II} \: \phi$ and $H_0 : \textrm{Class II} \: \phi >= \textrm{Class III}  \: \phi$.}
    \label{tab:class_comparison}
\end{table}

\section*{Discussion}

We present a simulation framework that couples a spatial agent-based model with an SIR epidemiological agent-based model to evaluate the relative risk reduction of intervention strategies towards the prevention of outbreaks of contagious respiratory diseases such as COVID-19. The spatial agent-based model, implemented in AnyLogic, used blueprints and building data to provide a high-fidelity representation of the room and hallway allotments of the modeled space, including constraints such as dedicated entrances and exits, restrooms, break areas, and arrival schedules. The spatial agent-based model was run under varying intervention strategies to generate contact networks. These intervention strategies included movement restrictions (one-way vs unrestricted), hourly break probability, and population size (relative to current recommended capacity). The resulting contact networks were in turn for simulations of the SIR model. Additionally, this two-stage modeling approach allowed us to cache the generated contact networks, which could then be used for further analysis through models other than the SIR model we have used in this study. This is especially important as the higher fidelity of the spatial agent-based model demands more computational resources and time.

Our experiments providing insights into which intervention strategies are more successful at mitigating outbreaks and why. Firstly, there is no significant reduction in risk by enforcing one-way movement (i.e. designating dedicated entrances or exits) over allowing for unrestricted movement. Instead, hourly break probability and population size have significant impacts on the risk of an outbreak, with the former having a much stronger effect. We confirm that this is due to the generation of contact networks with highly centralized hubs. The strongest impact on hub formation is seen with when individuals leave for breaks more often, followed by a lower, yet significant impact by population size. 

We identify three classes of interventions based on the possible final outbreak sizes (final infected ratios) produced by these two parameters, over all possible contagiousness values: Class I interventions with very strict restrictions on the frequency of leaving for breaks paired with very low population size, Class II interventions with a trade off between high (or low) break frequencies and low (or high) population size, and Class III interventions with both higher break frequencies and population size. We find that Class I restrictions lead to a significant decrease in outbreak size over Class II restrictions, for MERS, SARS, COVID-19, and Measles. Similarly, Class II restrictions lead to a significant decrease in outbreak size over Class III restrictions for all four diseases. Despite this advantage of Class I restrictions, it is important to consider the feasibility and ergonomic costs such recommendations have on personnel. When the hourly likelihood of leaving one's workstation is $0.05 \geq \alpha \leq 0.1$ this translates to an overall probability between $0.34$ and $0.57$ of taking at least one break during the whole 8 hour work day. Although individuals could still take breaks in between work inside their office spaces, this can lead to work-induced fatigue and isolation. In contrast, under Class II recommendations $0.25 \geq \alpha \leq 0.45$, leading to a probability between $0.90$ and $0.99$ that individuals leave their office space to take at least one 15 minute break within the floor or .. minutes outside the floor. However, Class II recommendations with $\alpha > 0.35$ require a population of $\beta=0.25$, a quarter of the recommended office capacity. In other words, despite the high efficacy of strict, Class I recommendations, a slightly more lenient Class II recommendation might be more feasible by reducing the population present below capacity and allowing individuals a much more generous flexibility of movement outside their designated workstations. 

This work provides a basis for future simulation studies that may benefit from the two-stage simulation approach presented in this paper. By modularizing the spatial and epidemiological aspects of contagious disease, we allow administrators and operations personnel to evaluate effects of spatial interventions independent of epidemiological model specifics made by computational epidemiologists. In other words, the spatial agent-based model may be thought of as a contact network generator for epidemiological models. Although, we have considered CDC definitions of \textit{prolonged contacts} in this study, minimal duration and distance of prolonged contacts can be treated as parameters to generate the respective contact networks. Extensions to the current spatial model could also include factors such as air-flow and ventilation to produce airborne transmission networks.

\bibliography{COVID19.bib}
\section*{Author contributions statement}


All authors assisted with data acquisition, and conceiving experiments, CG and RR conducted experiments, CG analyzed the results, and all authors reviewed the manuscript. 

\section*{Additional information}


\textbf{Competing interests} The authors declare no competing interests.


\end{document}